# Revealing the role of tin fluoride additive in narrow bandgap Pb-Sn perovskites for highly efficient flexible all-perovskite tandem cells

Johnpaul K. Pious[1],[#] Yannick Zwirner[1],[#] Huagui Lai[1],[#] Selina Olthof[2], Quentin Jeangros[3,4], Evgeniia Gilshtein[1], Radha K. Kothandaraman[1], Kerem Artuk[3], Philipp Wechsler[1], Cong Chen[5],[*] Christian M. Wolff[3], Dewei Zhao[5], Ayodhya. N. Tiwari[1], and Fan Fu[1*]

[1] Laboratory for Thin Films and Photovoltaics, Empa - Swiss Federal Laboratories for Materials Science and Technology, Dübendorf, Switzerland

[2] Department of Chemistry, University of Cologne, Greinstrasse 4–6, 50939, Cologne, Germany

[3] Photovoltaics and Thin Film Electronics Laboratory, EPFL - École Polytechnique Fédérale de Lausanne, Neuchâtel, Switzerland

[4] Sustainable Energy Center, Centre Suisse d'Electronique et de Microtechnique (CSEM), Jaquet-Droz 1, 2002 Neuchâtel, Switzerland

[5] College of Materials Science and Engineering, Engineering Research Center of Alternative Energy Materials & Devices, Ministry of Education, Sichuan University, Chengdu, Sichuan, 610065 China

[#] J.K.P., Y.Z., and H.L. contributed equally to this work.

[*] fan.fu@empa.ch; chen.cong@scu.edu.cn

**Abstract**

Tin fluoride ($SnF_2$) is an indispensable additive for high-efficiency Pb-Sn perovskite solar cells (PSCs). However, the spatial distribution of $SnF_2$ in the perovskite absorber is seldom investigated while essential for a comprehensive understanding of the exact role of the $SnF_2$ additive. Herein, we revealed the spatial distribution of $SnF_2$ additive and made structure-optoelectronic properties-flexible photovoltaic performance correlation. We observed the chemical transformation of $SnF_2$ to a fluorinated oxy-phase on the Pb-Sn perovskite film surface, due to its rapid oxidation. In addition, at the buried perovskite interface, we detected and visualized the accumulation of $F^-$ ions. We found that the photoluminescence quantum yield of Pb-Sn perovskite reached the highest value with 10 mol% $SnF_2$ in the precursor solution. When integrating the optimized absorber in flexible devices, we obtained the flexible Pb-Sn perovskite narrow bandgap (1.24 eV) solar cells with an efficiency of 18.5% and demonstrated 23.1%-efficient flexible 4-terminal all-perovskite tandem cells.

## Introduction

Solar cells based on perovskites with a general formula ABX$_3$ (A = methylammonium (MA$^+$), formamidinium (FA$^+$), and Cs$^+$; B = Pb$^{2+}$, Sn$^{2+}$; X = Cl$^-$, Br$^-$, and I$^-$) have shown rapid performance advancement, now reaching a certified power conversion efficiency (PCE) of 25.7%.[1] Further improvement in the efficiency of single junction perovskite solar cells (PSCs) will be restricted by the Shockley-Queisser limit.[2] Stacking two perovskite absorbers with complementary bandgaps in all-perovskite tandem solar cells (TSCs) holds great promise to go beyond the Shockley-Queisser efficiency limit of single-junction PSCs.[3-5]

Flexible all-perovskite TSC is an emerging field of research owing to its flexible and lightweight attributes, making them attractive for wearable electronics, avionics, building- and mobility-integrated photovoltaics.[6,7] In addition, the low-temperature solution processability of perovskites makes it viable for high throughput roll-to-roll (R2R) manufacturing, with the potential to substantially reduce the production cost and CO$_2$ footprint. Despite that, little progress has been made in this field, with only two publications reporting PCE of 21.3% and 24.4% for two-terminal (2T) flexible all-perovskite TSCs.[8,9] These values are well below the highest reported values for their rigid counterparts on 2-terminal (2T, 28.0%) and four-terminal (4T, 25.4%) tandem configuration, respectively.[10,11] So far there is no report on 4T all-perovskite tandem cells based on flexible substrates. Flexible foils pose several obstacles due to low thermal tolerance, high roughness, and poor wettability. Apart from that, it has proven difficult to grow homogenous narrow bandgap Pb-Sn perovskite films on flexible substrates owing to their fast crystallization.[12,13] Sn vacancy defect formation and its detrimental consequences in Pb-Sn perovskites are cause for concern, which becomes more severe when the concentration of Sn exceeds 50 mol% in the metal site.[14-18]

Tin fluoride (SnF$_2$) is an indispensable additive in the Pb-Sn perovskite precursor solution to suppress the oxidation of Sn$^{2+}$ ions. Previously, various groups have added different amounts of SnF$_2$ (5 mol% - 30 mol% relative to SnI$_2$) to fabricate Pb-Sn PSCs, but there is no consensus as to the optimal amount of SnF$_2$ needed, nor was the spatial distribution comprehensively investigated.[19-22] Furthermore, there are several competing explanations detailing the effects of SnF$_2$ as an additive on the crystal structure and local chemical environment in Pb-Sn perovskites. For instance, Zhao et al. proposed that SnF$_2$ fills the Sn vacancies and expands the perovskite lattice, which subsequently shifts the XRD peaks to a lower angle.[23] On the contrary, Herz and co-workers observed a lattice shrinkage and associated XRD peak shift to higher 2θ upon increasing the SnF$_2$ concentration; which was attributed to the Sn doping in the perovskite lattice.[24] The aforementioned reports imply that the spatial location or distribution of SnF$_2$ in the Pb-Sn perovskite film is not yet fully understood. In-depth understanding of the role of SnF$_2$ on the phase-composition, microstructure, and optoelectronic properties of Pb-Sn perovskite thin films as well as its spatial distribution require further investigation to improve the efficiency and stability of Pb-Sn PSCs.

In this work, we varied the SnF$_2$ concentration added to the (FASnI$_3$)$_{0.6}$(MAPbI$_3$)$_{0.4}$ perovskite precursor to systematically investigate its effects on the film morphology, crystal structure, optoelectronic properties, and flexible solar cell device performance. We unraveled that SnF$_2$ addition modulates the perovskite grain topography and suppresses the defect formation via a p-type self-doping mechanism. SnF$_2$ segregates both at the hole transport layer/perovskite and electron transport layer/perovskite interfaces. We observe the formation of fluorinated tin dioxide at the top surface of the (FASnI$_3$)$_{0.6}$(MAPbI$_3$)$_{0.4}$ films. Although SnF$_2$ improved the optoelectronic quality of the Pb-Sn absorber, beyond a threshold of 10% further addition of SnF$_2$ impairs the device performance through a lowering of the FF and $V_{OC}$. As a result of the optimized SnF$_2$ supplementation, we achieved flexible narrow bandgap PSC with a champion efficiency of 18.5%. Furthermore, we demonstrated over 23% flexible 4-terminal all perovskite tandem solar cell combining a 1.24 eV Pb-Sn bottom subcell and a 1.78 eV perovskite top cell.

**Results and Discussion**

(FASnI$_3$)$_{0.6}$(MAPbI$_3$)$_{0.4}$ perovskite thin films with different mol% of SnF$_2$ additive in the precursor solution were deposited using an anti-solvent assisted method as described in the experimental section. The scanning electron microscopy (SEM) images of perovskite films deposited on PEN/ITO/PEDOT:PSS substrate showed smooth morphology up to 2.5mol% SnF$_2$ (**Figures** 1a and b), while a further increase in concentration resulted in surface texture. For samples with 5mol% SnF$_2$, terraces-like structures start to appear as shown in the SEM (**Figure** 1c) and atomic force microscopy (AFM) images (**Figure** S1). These films exhibited a high root mean square roughness ($R_q$) value of 36.7 compared to control samples without SnF$_2$ ($R_q$ = 29.6).

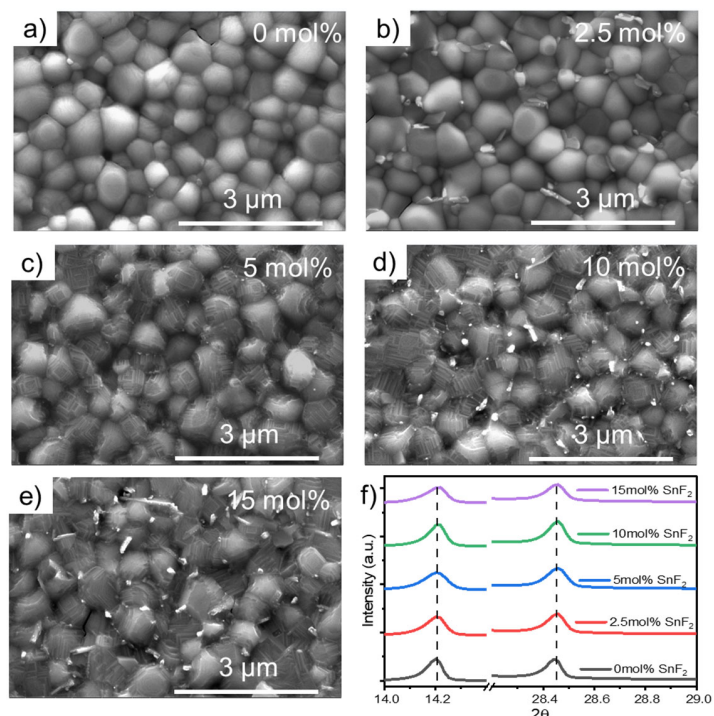

**Figure 1**. a) – e) SEM surface morphology of Pb-Sn perovskite films with different mol% of SnF$_2$ additive. f) XRD patterns of the perovskite films with various amounts of SnF$_2$.

Thin films with 10mol% and 15mol% SnF$_2$ content exhibited "white spots" in the SEM images (**Figures** 1d and e). We believe that the textured features on the perovskite grain surfaces might have originated from the modulated colloidal property of the perovskite precursor solution. This could be ascribed to the [SnI$_x$]$^{2-x}$ adduct formation as a result of SnF$_2$ complexation with Sn$^{4+}$ ions as reported previously.[25]

To understand the role of SnF$_2$ on the crystal structure of the Pb-Sn perovskite, the XRD patterns of the perovskite films with various amounts of SnF$_2$ were analyzed. Preferential orientation of Pb-Sn perovskite crystallites along the (110) and (220) lattice planes was observed (**Figure** 1f). The perovskite films retained a similar crystallinity up to 15mol% SnF$_2$ addition as indicated by their similar diffraction peak intensity. There was no noticeable XRD peak shift upon increasing the SnF$_2$ concentration up to 15 mol%. Previous reports proposing the Sn vacancy filling and metal site doping during SnF$_2$ supplementation showed XRD peak shifts to lower or higher 2θ values, respectively.[19, 23] However, our XRD measurements suggest that SnF$_2$ does not alter the perovskite lattice.

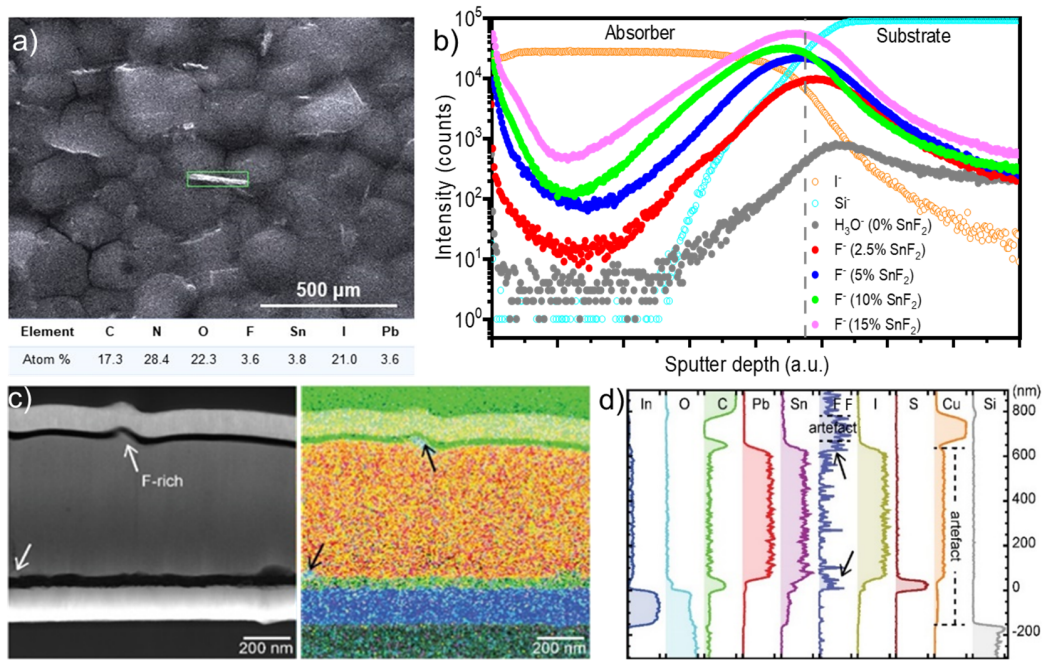

**Figure 2**. a) Scanning electron microscopy (SEM) image and energy dispersive X-ray (EDX) elemental analysis of (FASnI$_3$)$_{0.6}$(MAPbI$_3$)$_{0.4}$ thin film with 10mol% SnF$_2$. The chemical composition in the area of white aggregate (green rectangular box) is shown in the table. b) ToF-SIMS depth profiles of (FASnI$_3$)$_{0.6}$(MAPbI$_3$)$_{0.4}$ perovskite thin films with different concentrations of SnF$_2$. c) Cross-sectional scanning transmission electron microscopy high-angle annular dark-field (STEM HAADF) image with the corresponding EDX mapping and (d) EDX line profiles for the flexible Pb-Sn perovskite solar cell. Arrows indicate the fluoride-rich regions at the top and bottom interfaces of perovskite with transport layers.

To investigate the composition of the white spots observed in the SEM images, we performed an EDX analysis and identified the presence of fluorine, which hinted at the presence of $SnF_2$ (**Figure** 2 a). Even though it is important to understand the distribution of $SnF_2$ in the perovskite film to explicitly unravel the role of $SnF_2$ additive, so far only a little research attention was paid in this direction. To shed light on this missing knowledge, we performed the time-of-flight secondary ion mass spectrometry (ToF-SIMS) depth profiling measurements on perovskite/glass layer stacks. The marker $F^-$ secondary ion signal representing $SnF_2$ was detected throughout the film, with a higher signal strength both at the top and bottom interfaces of the perovskite absorber, irrespective of the concentration. At low concentrations of $SnF_2$ (2.5 mol% and 5 mol%), $F^-$ ions appear to preferentially segregate at the buried perovskite/HTL interface. At higher $SnF_2$ supplementation, the top surface of the perovskite layers has shown $F^-$ ion signal intensity on par with the bottom interface (**Figure** 2 b). Generally, the summed $F^-$ ion signal throughout the layer increases with an increase in the amount of $SnF_2$ in the perovskite precursor solution. To further visualize the spatial distribution of $SnF_2$, we conducted cross-sectional EDX mapping using HAADF-STEM. The EDX mapping of full device stacks with 10mol% $SnF_2$ confirms the accumulation of $SnF_2$ both at the perovskite/HTL and perovskite/ETL interfaces (**Figure** 2 c).

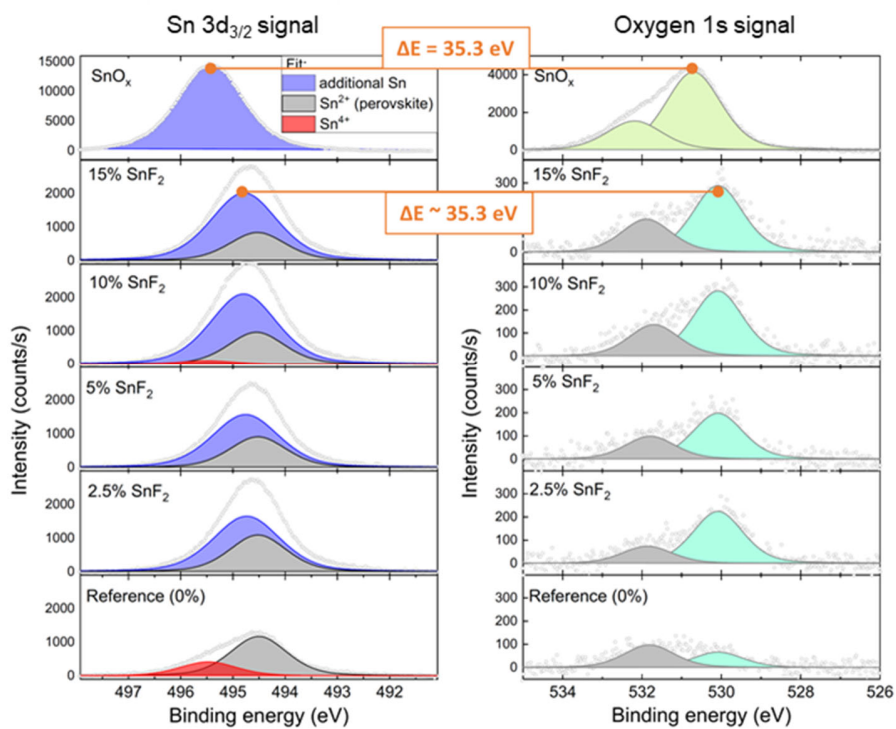

**Figure 3**. XPS core level spectra of $Sn3d_{3/2}$ and $O1s$ of $SnO_2$ as well as the Pb-Sn perovskite films with varied amounts of $SnF_2$ additive.

X-ray photoelectron spectroscopy (XPS) was performed to study in detail the composition of the surfaces after the addition of different amounts of $SnF_2$ and to analyze the element oxidation states. The $Sn3d_{3/2}$ signal is shown on the left side of **Figure** 3. Fitting this feature is rather challenging since Sn can show up at different binding energies. Possible chemical environments are intact perovskite ($Sn^{+2}$), degraded $Sn^{+4}$ phases, intact $SnF_2$ domains, or even new species, as will be revealed below. The rigorous fitting procedure is detailed in the Supporting Information and the resulting features are indicated in **Figure** 3. For the perovskite film without SnF$_2$ additive, a clear signal originating from Sn$^{+4}$ is seen as marked by the red peak at higher binding energy, in addition to the perovskite-related signal (grey). Once SnF$_2$ is added, another rather intense Sn signal appears, marked in blue, with binding energy in between the ones observed for Sn$^{+2}$ and Sn$^{+4}$. This additional peak makes overlaps with the Sn$^{4+}$ signal, reducing the ability to quantify the latter. Even though the fits are less robust, our analysis indicates that the amount of Sn$^{+4}$ is likely negligible in the SnF$_2$-containing samples.

Also, the F1s core level signal was analyzed upon adding SnF$_2$ to the perovskite precursor solution, which confirms the presence of fluorine on the surface of the perovskite layer (**Figure** S2). To determine whether the fluorine is still bound to Sn, we analyzed also a pure SnF$_2$ film which is included in **Figure** S2. Notably, the binding energy difference between F1s and Sn3d$_{3/2}$ signals in pure SnF$_2$ ($\Delta E=188.67$ eV) does not match the one observed for F1s and the newly emerged blue Sn feature ($\Delta E=189.29$ eV). This suggests the formation of a new chemical species. A second indication that the additional Sn is not (exclusively) bound to fluorine comes from the analysis of the relative concertation of the elements which is SnF$_x$ (x = 0.2 – 0.5), meaning that there is a severe lack of F signal.

Interestingly, SnF$_2$ supplementation resulted in the appearance of an oxygen O1s signal at lower binding energy with high intensity compared to the reference sample without SnF$_2$, see Figure 3 on the right. From these observations, we hypothesized the new species formed to be fluorinated SnO$_2$ because SnF$_2$ on the exposed perovskite surface can be readily oxidized in presence of oxygen and hence serve as an oxygen scavenger to hinder the degradation of the absorber. To further confirm this, we compared the binding energy difference ($\Delta E$) of Sn (3d$_{3/2}$) and O (1s) signals from a pure SnO$_2$ sample (**Figure** 3 top) with that in the Pb-Sn perovskite films. The Sn3d$_{3/2}$ vs.O1s core level difference in this SnO$_2$ sample is $\Delta E = 35.28$ eV, which matches with the ones observed in the perovskite absorbers ($\Delta E = 35.31 \pm 0.03$ eV) validating the formation of fluorinated SnO$_2$ on the surface of the perovskite layer. From the fitted XPS data of the perovskite films with different concentrations of SnF$_2$, we estimated the composition of the newly formed inorganic species as SnO$_y$F$_x$ (x = 0.2 – 0.5, y = 0.9 - 1.7).

The effect of the SnF$_2$ additive on the optoelectronic quality of Pb-Sn perovskites supplemented with different amounts of SnF$_2$ was examined using photoluminescence quantum yield (PLQY) measurements (**Figure** 4a). Since the charge extraction layers are known to introduce additional nonradiative recombination, glass/PEDOT:PSS/(FASnI$_3$)$_{0.6}$(MAPbI$_3$)$_{0.4}$/C$_{60}$ stacks were used for the PLQY measurements to ensure a reliable correlation with the actual device performance. The PLQY values showed an increasing trend up to 10 mol% SnF$_2$ addition and then decreased at higher SnF$_2$ concentration (15 mol%). This trend could be attributed to the improved suppression of Sn oxidation up to 10 mol% SnF$_2$ addition in the precursor solution, and then roughly stagnates. This is in line with a reduction of nonradiative bulk recombination due to reduced oxidation during and after crystallization and in addition a reduction of interfacial recombination due to the formation of F-enriched phases at both interfaces. Therefore, the optimal concentration of SnF$_2$ additive to minimize the non-radiative recombination in (FASnI$_3$)$_{0.6}$(MAPbI$_3$)$_{0.4}$ based PSC devices would be 10 mol%. In line with the above observation, ultraviolet photoelectron spectroscopy analysis showed a notable suppression of states close to the Fermi level when $SnF_2$ was introduced into the perovskite films; these are likely related to gap states tailing into the band gap (**Figure** S3). This could be attributed to the reduction of detrimental p-type self-doping in Pb-Sn perovskite films supplemented with $SnF_2$. To verify the hypothesis of $SnF_2$-mediated defect healing and to investigate the role of $SnF_2$ on PV performance, we have fabricated Pb-Sn (NBG) PSCs on flexible substrates.

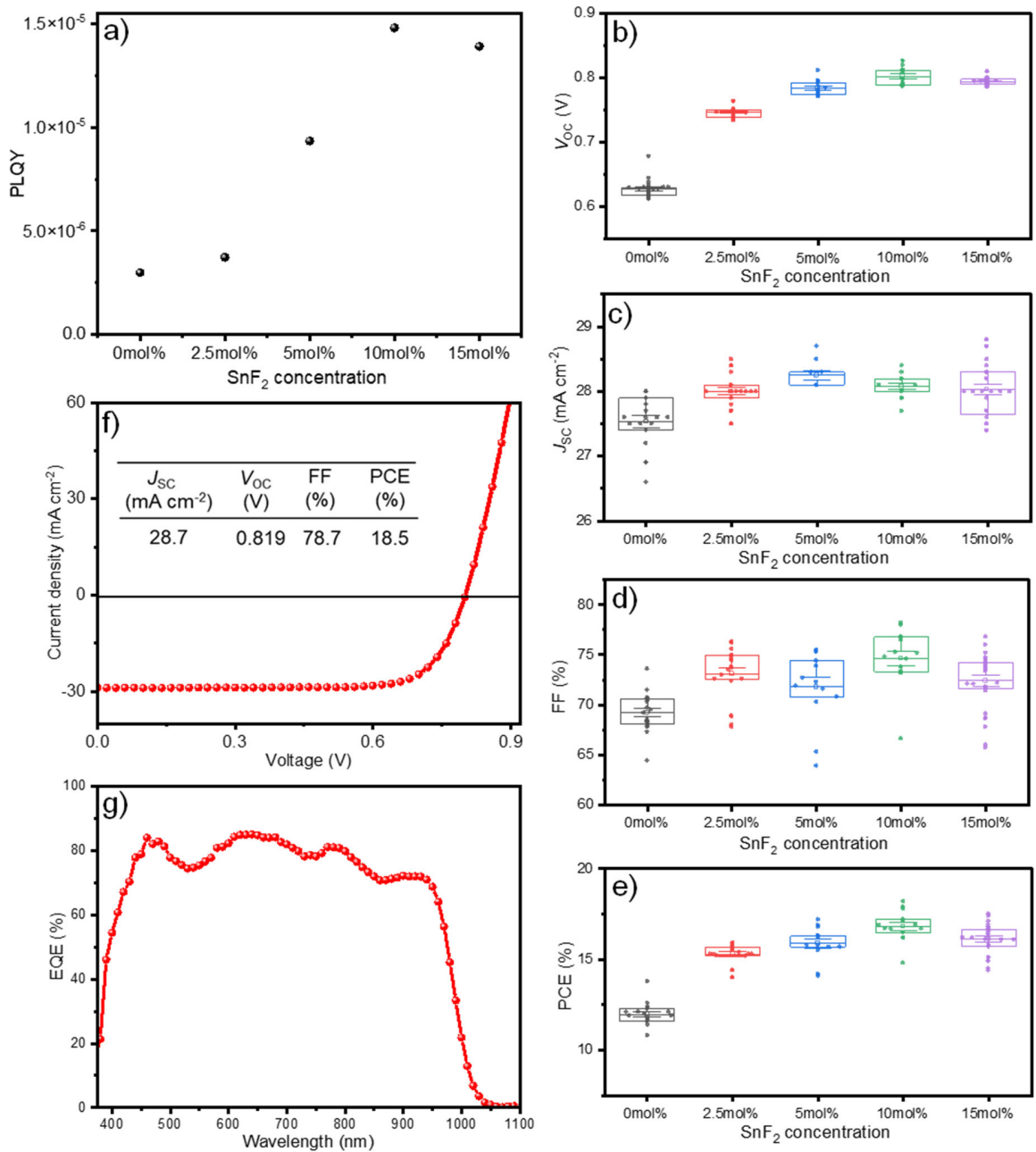

**Figure 4**. a) Photoluminescence quantum yield of the perovskite absorber with different amounts of $SnF_2$ additive sandwiched between the charge extraction layers (HTL and ETL). The box chart shows the variation in Pb-Sn PSC parameters such as b) $V_{OC}$, c) $J_{SC}$, d) FF, and e) PCE with different mol% of SnF$_2$ additive. f) and g) are *J-V* and EQE plots of the champion flexible NBG device.

The *J-V* parameters of the flexible NBG PSCs fabricated with different SnF$_2$ additive concentrations are summarized in **Figure** 4b-e. The box plot shows a clear trend of open-circuit voltage (*V*$_{OC}$) improvement upon increasing the amount of SnF$_2$ up to 10 mol% reaching a maximum average value of 810 mV. For higher SnF$_2$ concentrations (15 mol%), the average value of *V*$_{OC}$ is slightly decreased (**Figure** 4b). The average value of fill factor (FF) increased up to 10mol% SnF$_2$ supplementation, implying an improved intimate contact between perovskite and charge selective contacts (**Figure** 4d). The drop in the average value of FF above 10 mol% SnF$_2$ might be due to the excess SnF$_2$ aggregation at the front and rear interfaces of the perovskite absorber. The *V*$_{OC}$ of (FASnI$_3$)$_{0.6}$(MAPbI$_3$)$_{0.4}$ based devices showed a similar trend (**Figure** 4e) as that of the PLQY data, implying that the SnF$_2$ additive improves the optoelectronic quality of the absorber, but also the whole stack. As a result, the flexible NBG PSC with 10 mol% SnF$_2$ exhibited the optimal performance with a champion efficiency of 18.5%, corresponding to a *V*$_{OC}$ of 0.819 V, a *J*$_{SC}$ of 28.7 mA cm$^{-2}$, and FF of 78.7% (**Figure** 4f). The device delivered a steady-state efficiency of 18.2% during maximum power point tracking (**Figure** S4). The integrated *J*$_{SC}$ value of 28.5 mA cm$^{-2}$ obtained from the EQE spectrum (**Figure** 4g) is in good agreement with the *J*$_{SC}$ value determined from the *J-V* measurement.

To verify the potential of our flexible Pb-Sn PSCs in the application of all-perovskite TSCs, we mechanically stacked the champion flexible Pb-Sn PSC with the flexible wide bandgap PSC developed in our lab and demonstrate a 4T flexible all-perovskite TSC as shown in **Figure** 5a. The best performing flexible 4T all-perovskite TSC fabricated using a 1.24 eV Pb-Sn bottom subcell and a 1.78 eV Cs$_{0.12}$FA$_{0.8}$MA$_{0.08}$PbI$_{1.8}$Br$_{1.2}$ top subcell delivered a PCE of 23.1% (**Figure** 5b). The integrated *J*$_{SC}$ values obtained from EQE spectra for the WBG top cell and NBG bottom cell are 15.46 and 12.45 mA cm$^{-2}$, respectively (**Figure** 5c). To the best of our knowledge, this is the first report of flexible Pb-Sn NBG (1.24 eV) single junction and flexible 4T all-perovskite TSCs (**Figure** 5d).[8,9,26]

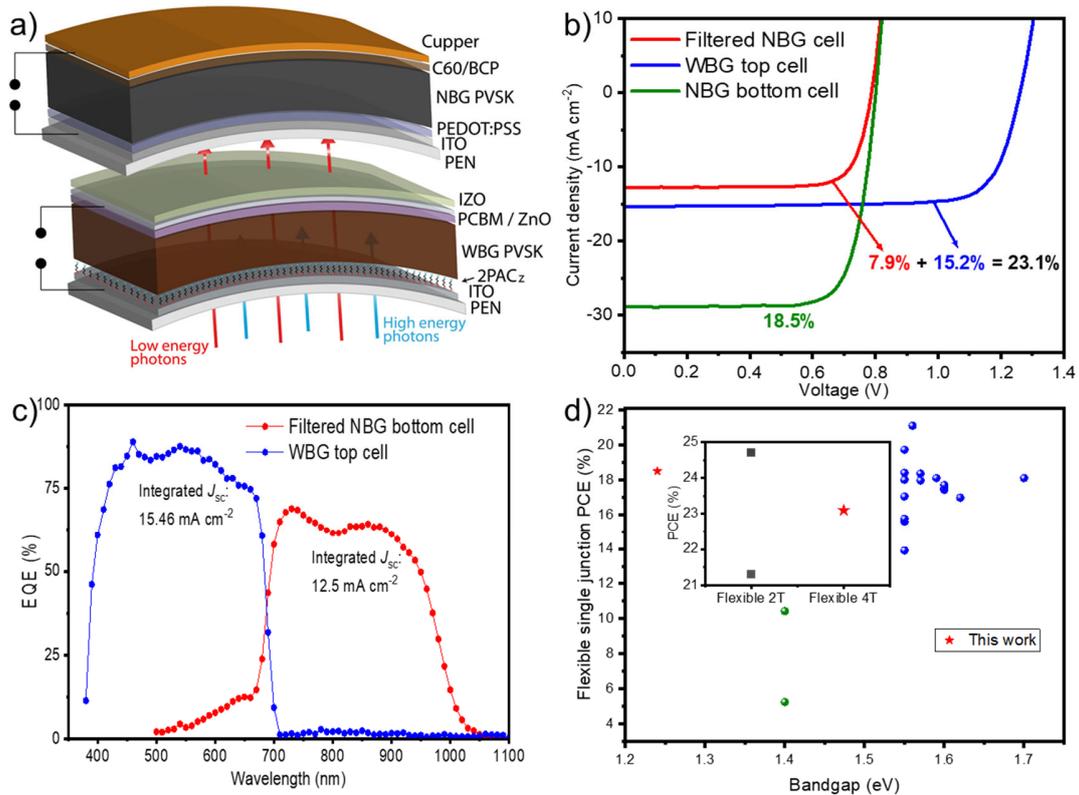

**Figure 5**. a) Schematic of a flexible 4T all-perovskite TSC. b) and c) are *J-V* and EQE characteristics of individual subcells in the 4T tandem configuration. d) Plot of reported flexible single junction PSCs efficiencies versus bandgap. The inset graph shows the reported flexible all-perovskite TSC efficiencies. Red asterisks represent the PCE values of our devices presented in this work.

**Conclusion**

In conclusion, we unraveled that the addition of $SnF_2$ to the precursor ink induces a textured morphology in the Pb-Sn perovskite films, and the $F^-$ ions from $SnF_2$ segregate both at the top and bottom interfaces of the perovskite layer. Employing STEM, we have visualized the presence of $F^-$ ions islands at the perovskite/HTL and perovskite/ETL interfaces. We observed that the $SnF_2$ present on the exposed perovskite surface is easily oxidized resulting in the formation of $SnO_{1.2}F_{(0.2-0.5)}$. Although $SnF_2$ doesn't have any noticeable impact on the perovskite crystal structure, electronic defects at the valence band edge states are reduced, reducing recombination and hence allowing an increase in the quasi Femi-level splitting of the absorber. Excess $SnF_2$ does not further improve the $V_{OC}$ of the NBG PSCs, but results in a mildly reduced fill factor and performance. By carefully optimizing the concentration of $SnF_2$, we achieved a flexible NBG single-junction PSC efficiency of up to 18.5% and a 4-T tandem efficiency of 23.1%. Our comprehensive experimental investigation including surface and depth elemental profiling techniques in combination with the device performance evaluation unveiled the multiple unknown roles of $SnF_2$ additive and a new fluorinated tin oxide ($SnO_{1.2}F_{(0.2-0.5)}$) based interface formation in Pb-Sn PSCs.

ASSOCIATED CONTENT

**Supporting Information**

Experimental details, AFM images, XPS and UPS spectra, and MPP stability tracking data.

AUTHOR INFORMATION

**Notes**

The authors declare no competing financial interest.

**Acknowledgments**
This work was received funding from the Swiss National Science Foundation (N. 200021_213073/1) and European Union's Horizon Europe research and innovation programme under grant agreement No 101075605. The authors acknowledge the financial support from the Empa internal call 2021 (TexTandem) and Strategic Focus Area Advanced Manufacturing under the project AMYS-Advancing manufacturability of hybrid organic-inorganic semiconductors for large-area optoelectronics. The work was financially supported by the National Key Research and Development Program of China (No. 2019YFE0120000), Fundamental Research Funds for the Central Universities (Nos. YJ2021157 and YJ201955), Engineering Featured Team Fund of Sichuan University (No. 2020SCUNG102). S.O. acknowledges funding by the German Federal Ministry for Education and Research (MUJUPO$^2$, Grant OL 462/4-2). C.M.W. thanks the European Commission for funding through Marie Skłodowska-Curie Actions (Grant No. 101033077). K.A., Q. J., and C.M.W. thank the Swiss National Science Foundation (PAPET, 197006) and the Swiss Federal Office of Energy (PRESTO) for funding. J.K.P. acknowledges the financial support from the Swiss Government Excellence Research Scholarship (ESKAS Ref No. 2021.0332). H.L. thanks the China Scholarship Council (CSC) funding from the Ministry of Education of P. R. China.

**References**
1. Best Research-Cell Efficiency Chart; https://www.nrel.gov/pv/cell-efficiency.html (2021), Accessed 8[th] June 2022
2. Kothandaraman, R. K.; Jiang, Y.; Feurer, T.; Tiwari, A. N.; Fu, F. Near-Infrared-Transparent Perovskite Solar Cells and Perovskite-Based Tandem Photovoltaics. *Small Methods* **2020**, *4*, 2000395.
3. Leijtens, T.; Bush, K. A.; Prasanna, R.; McGehee, M. D. Opportunities and challenges for tandem solar cells using metal halide perovskite semiconductors. *Nat. Energy* **2018** 828–838.
4. Wang, Y.; Zhang, M.; Xiao, K.; Lin, R.; Luo, X.; Han, Q.; Tan, H. Recent progress in developing efficient monolithic all-perovskite tandem solar cells. *J. Semicond.* **2020**, *41*, 051201.

5. Zheng, X.; Alsalloum, A. Y.; Hou, Y.; Sargent, E. H.; Bakr, O. M. All-Perovskite Tandem Solar Cells: A Roadmap to Uniting High Efficiency with High Stability. *Acc. Mater. Res.* **2020**, *1*, 63−76.
6. Gao, Y.; Huang, K.; Long, C.; Ding, Y.; Chang, J.; Zhang, D.; Etgar, L.; Liu, M.; Zhang, J.; Yang, J. Flexible Perovskite Solar Cells: From Materials and Device Architectures to Applications. *ACS Energy Lett.* **2022**, *7*, 1412−1445.
7. Zhang, J.; Zhang, W.; Cheng, H.-M.; Silva, S. R. P. Critical Review of Recent Progress of Flexible Perovskite Solar Cells. *Mater. Today* **2020**, *39*, 66−88.
8. Palmstrom, A. F.; Eperon, G. E.; Leijtens, T.; Prasanna, R.; Habisreutinger, S. N.; Nemeth, W.; Gaulding, E. A.; Dunfield, S. P.; Reese, M.; Nanayakkara, S.; Moot, T.; Werner, J. M.; Liu, J.; To, B.; Christensen, S. T.; McGehee, M. D.; van Hest, M. F.A.M.; Luther, J. M.; Berry, J. J.; Moore, D. T. Enabling Flexible All-Perovskite Tandem Solar Cells. *Joule* **2019**, 2193–2204.
9. Li, L.; Wang, Y.; Wang, X.; Lin, R.; Luo, X.; Liu, Z.; Zhou, K.; Xiong, S.; Bao, Q.; Chen, G.; Tian, Y.; Deng, Y.; Xiao, K.; Wu, J.; Saidaminov, M. I.; Lin, H.; Ma, C.-Q.; Zhao, Z.; Wu, Y.; Zhang, L.; Tan, H. Flexible all-perovskite tandem solar cells approaching 25% efficiency with molecule-bridged hole-selective contact. *Nat. Energy* **2022**, https://doi.org/10.1038/s41560-022-01045-2.
10. Tong, J.; Song, Z.; Kim, D. H.; Chen, X.; Chen, C.; Palmstrom, A. F.; Ndione, P. F.; Reese, M. O.; Dunfield, S. P.; Reid, O. G.; Liu, J.; Zhang, F.; Harvey, S. P.; Li, Z.; Christensen, S. T.; Teeter, G.; Zhao, D.; Al-Jassim, M. M.; van Hest, M. F. A. M.; Beard, M. C.; Shaheen, S. E.; Berry, J. J.; Yan, Y.; Zhu, K. Carrier lifetimes of >1 μs in Sn-Pb perovskites enable efficient all-perovskite tandem solar cells. *Science* **2019**, *364*, 475–479.
11. Lin, R.; Xu, J.; Wei, M.; Wang, Y.; Qin, Z.; Liu, Z.; Wu, J.; Xiao, K.; Chen, B.; Park, S. M.; Chen, G.; Atapattu, H. R.; Graham, K. R.; Xu, J.; Zhu, J.; Li, L.; Zhang, C.; Sargent, E. H.; Tan, H. All-perovskite tandem solar cells with improved grain surface passivation. *Nature* **2022**, *603*, 73–78.
12. Hu, Y.; Niu, T.; Liu, Y.; Zhou, Y.; Xia, Y.; Ran, C.; Wu, Z.; Song, L.; Müller- Buschbaum, P. Chen, Y.; Huang, W. Flexible Perovskite Solar Cells with High Power-Per-Weight: Progress, Application, and Perspectives. *ACS Energy Lett.* **2021**, *6*, 2917−2943.
13. Liu, H.; Wang, L.; Li, R.; Shi, B.; Wang, P.; Zhao, Y.; Zhang, X. Modulated Crystallization and Reduced $V_{OC}$ Deficit of Mixed Lead−Tin Perovskite Solar Cells with Antioxidant Caffeic Acid. *ACS Energy Lett.* **2021**, *6*, 2907−2916.
14. Ghimire, N.; Bobba, R. S.; Gurung, A.; Reza, K. M.; Laskar, M. A. R.; Lamsal, B. S.; Emshadi, K.; Pathak, R.; Afroz, M. A.; Chowdhury, A. H.; Chen, K.; Bahrami, B.; Rahman, S. I.; Pokharel, J.; Baniya, A.; Rahman, M. T.; Zhou, Y.; Qiao, Q. Mitigating Open-Circuit Voltage Loss in Pb−Sn Low-Bandgap Perovskite Solar Cells via Additive Engineering. *ACS Appl. Energy Mater.* **2021**, *4*, 1731−1742.
15. Kapil, G.; Bessho, T.; Sanehira, Y.; Sahamir, S. R.; Chen, M.; Baranwal, A. K.; Liu, D.; Sono, Y.; Hirotani, D.; Nomura, D.; Nishimura, K.; Kamarudin, M. A.; Shen, Q.; Segawa, H.; Hayase, S. Tin-Lead Perovskite Solar Cells Fabricated on Hole Selective Monolayers. *ACS Energy Lett.* **2022**, *7*, 966−974.

16. Lin, R.; Xiao, K.; Qin, Z.; Han, Q.; Zhang, C.; Wei, M.; Saidaminov, M. I.; Gao, Y.; Xu, J.; Xiao, M.; Li, A.; Zhu, J.; Sargent, E. H.; Tan, H. Monolithic all-perovskite tandem solar cells with 24.8% efficiency exploiting comproportionation to suppress Sn(II) oxidation in precursor ink. *Nat. Energy* **2019**, *4*, 864–873.
17. Xiao, K.; Lin, R.; Han, Q.; Hou, Y.; Qin, Z.; Nguyen, H. T.; Wen, J.; Wei, M.; Yeddu,V.; Saidaminov, M. I.; Gao, Y.; Luo, X.; Wang, Y.; Gao, H.; Zhang, C.; Xu, J.; Zhu, J.; Sargent, E. H.; Tan, H. All-perovskite tandem solar cells with 24.2% certified efficiency and area over 1 cm$^2$ using surface-anchoring zwitterionic antioxidant. *Nat. Energy* **2020**, *5*, 870–880.
18. Hu, S.; Otsuka, K.; Murdey, R.; Nakamura, T.; Truong, M. A.; Yamada, T.; Handa, T.; Matsuda, K.; Nakano, K.; Sato, A.; Marumoto, K.; Tajima, K.; Kanemitsu, Y.; Wakamiya, A. Optimized Carrier Extraction at Interfaces for 23.6% Efficient Tin–Lead Perovskite Solar Cells. *Energy Environ. Sci*. **2022**, *15*, 2096–2107.
19. Ke, W.; Chen, C.; Spanopoulos, I.; Mao, L.; Hadar, I.; Li, X.; Hoffman, J. M.; Song, Z.; Yan, Y.; Kanatzidis, M. G. Narrow-Bandgap Mixed Lead/Tin-Based 2D Dion-Jacobson Perovskites Boost the Performance of Solar Cells. *J. Am. Chem. Soc*. **2020**, *142*, 15049−15057.
20. Tsai, C.-M.; Wu, H.-P.; Chang, S.-T.; Huang, C.-F.; Wang, C.-H.; Narra, S.; Yang, Y.-W.; Wang, C.-L.; Hung, C.-H.; Diau, E. W.-G. Role of Tin Chloride in Tin-Rich Mixed-Halide Perovskites Applied as Mesoscopic Solar Cells with a Carbon Counter Electrode. *ACS Energy Lett*. **2016**, *1*, 1086–1093.
21. Rajagopal, A.; Liang, P.-W.; Chueh, C.-C.; Yang, Z.; Jen, A. K.-Y. Defect Passivation via a Graded Fullerene Heterojunction in Low-Bandgap Pb−Sn Binary Perovskite Photovoltaics. *ACS Energy Lett*. **2017**, *2*, 2531−2539.
22. Igual-Munoz, A. M.; Ávila, J.; Boix, P. P.; Bolink, H. J. FAPb$_{0.5}$Sn$_{0.5}$I$_3$: A Narrow Bandgap Perovskite Synthesized through Evaporation Methods for Solar Cell Applications. *Sol. RRL* **2020**, *4*, 1900283.
23. Chen, Q.; Luo, J.; He, R.; Lai, H.; Ren, S.; Jiang, Y.; Wan, Z.; Wang, W.; Hao, X.; Wang, Y.; Zhang, J.; Constantinou, I.; Wang, C.; Wu, L.; Fu, F.; Zhao, D. Unveiling Roles of Tin Fluoride Additives in High-Efficiency Low-Bandgap Mixed Tin-Lead Perovskite Solar Cells. *Adv. Energy Mater*. **2021**, *11*, 2101045.
24. Savill, K. J.; Ulatowski, A. M.; Farrar, M. D.; Johnston, M. B.; Snaith, H. J.; Herz, L. M. Impact of Tin Fluoride Additive on the Properties of Mixed Tin-Lead Iodide Perovskite Semiconductors. *Adv. Funct. Mater*. **2020**, *30*, 2005594.
25. Pascual, J.; Flatken, M.; Félix, R.; Li, G.; Turren-Cruz, S.-H.; Aldamasy, M. H.; Hartmann, C.; Li, M.; Girolamo, D. D.; Nasti, G.; Hgsam, E.; Wilks, R. G.; Dallmann, A.; Bär, M.; Hoell, A.; Abate, A. Fluoride Chemistry in Tin Halide Perovskites. *Angew. Chem. Int. Ed*. **2021**, *60*, 21583 – 21591.
26. Chung, J.; Shin, S. S.; Hwang, K.; Kim, G.; Kim, K. W.; Lee, D. S.; Kim, W.; Ma, B. S.; Kim, Y.-K.; Kim, T.-S.; Seo, J. Record-efficiency flexible perovskite solar cell and module enabled by a porous-planar structure as an electron transport layer. *Energy Environ. Sci*., **2020**, *13*, 4854-4861.